\documentclass[fleqn,usenatbib]{mnras}

\usepackage{newtxtext,newtxmath}
\usepackage[T1]{fontenc}
\usepackage{xcolor}
\usepackage{pdfpages}

\usepackage{float}
\usepackage{svg}

\svgsetup{inkscapelatex=true}
\usepackage{graphicx}

\usepackage{ragged2e}
\usepackage[utf8]{inputenc}
\usepackage{booktabs}

\usepackage[T1]{fontenc}
\usepackage{multirow}

\DeclareRobustCommand{\VAN}[3]{#2}
\let\VANthebibliography\thebibliography
\def\thebibliography{\DeclareRobustCommand{\VAN}[3]{##3}\VANthebibliography}

\usepackage{graphicx}	
\usepackage{amsmath}	
\usepackage{float}

\usepackage{url}
\usepackage{xcolor}

\title{Classification of Radio Galaxies with trainable COSFIRE filters}

\author[S.Ndung'u et al.]{
Steven Ndung'u,$^{1,2}$\thanks{E-mail: s.n.machetho@rug.nl}
Trienko Grobler,$^{2}$
Stefan J. Wijnholds$^{2,3}$
Dimka Karastoyanova$^{1}$
and George Azzopardi$^{1}$
\\
$^{1}$University of Groningen, Nijenborgh 9, 9712 CP, Groningen, The Netherlands\\
$^{2}$University of Stellenbosch, Cnr Banhoek Road \& Joubert Street, Stellenbosch, 7600, South Africa\\
$^{3}$ASTRON, Oude Hoogeveensedijk 4, 7991 PD. Dwingeloo, The Netherlands
}

\date{Accepted XXX. Received YYY; in original form ZZZ}

\pubyear{2023}

\begin{document}
\label{firstpage}
\pagerange{\pageref{firstpage}--\pageref{lastpage}}
\maketitle

\begin{abstract}

\noindent Radio galaxies exhibit a rich diversity of characteristics and emit radio emissions through a variety of radiation mechanisms, making their classification into distinct types based on morphology a complex challenge. To address this challenge effectively, we introduce an innovative approach for radio galaxy classification using COSFIRE filters. These filters possess the ability to adapt to both the shape and orientation of prototype patterns within images. The COSFIRE approach is explainable, learning-free, rotation-tolerant, efficient, and does not require a huge training set. To assess the efficacy of our method, we conducted experiments on a benchmark radio galaxy data set comprising of 1180 training samples and 404 test samples. Notably, our approach achieved an average accuracy rate of 93.36\%. This achievement outperforms contemporary deep learning models, and it is the best result ever achieved on this data set. Additionally, COSFIRE filters offer better computational performance, $\sim$20$\times$ fewer operations than the DenseNet-based competing method (when comparing at the same accuracy). Our findings underscore the effectiveness of the COSFIRE filter-based approach in addressing the complexities associated with radio galaxy classification. This research contributes to advancing the field by offering a robust solution that transcends the orientation challenges intrinsic to radio galaxy observations. Our method is versatile in that it is applicable to various image classification approaches.

\end{abstract}




\section{Introduction}

Automatically classifying radio galaxies according to their morphology has been extensively studied within the literature over the last few years. This has been motivated by the introduction of various innovative and diverse machine/deep learning techniques. Three main factors have driven the adoption of these state-of-the-art methods in the processing of interferometric data and images: the availability of high-resolution images from modern radio telescopes such as MeerKAT \citep{knowles2022meerkat} and LOFAR \citep{shimwell2022lofar}; the availability of labelled data sets from public initiatives such as Radio Galaxy Zoo \citep{banfield2015radio}, LOFAR Galaxy Zoo\footnote{https://www.zooniverse.org/projects/chrismrp/radio-galaxy-zoo-lofar}, and other catalogs \citep{proctor2011morphological,10.1111/j.1365-2966.2012.20414.x,Baldi2015APS,10.1093/mnras/stx007,capetti2017fricat,capetti2017friicat,baldi2018fr0cat,ma2019machine}; and the interdisciplinary and collaborative research with other fields such as computer science, statistics, and data science. These techniques have opened up new possibilities in the field of radio astronomy, as they provide fundamental insights into astrophysical phenomena and facilitate serendipitous discoveries. Automatic galaxy classification is essential for understanding the physical processes that shape and transform radio galaxies \citep{ndungu2023advances}. This is especially true in the case of radio interferometry, due to the large data sets that are currently being generated by modern radio surveys. These data sets can be terabytes to exabytes in size \citep{booth2012overview,10.1117/1.JATIS.8.1.011021}.

\begin{figure}
 \centering
\includegraphics[width=\columnwidth]{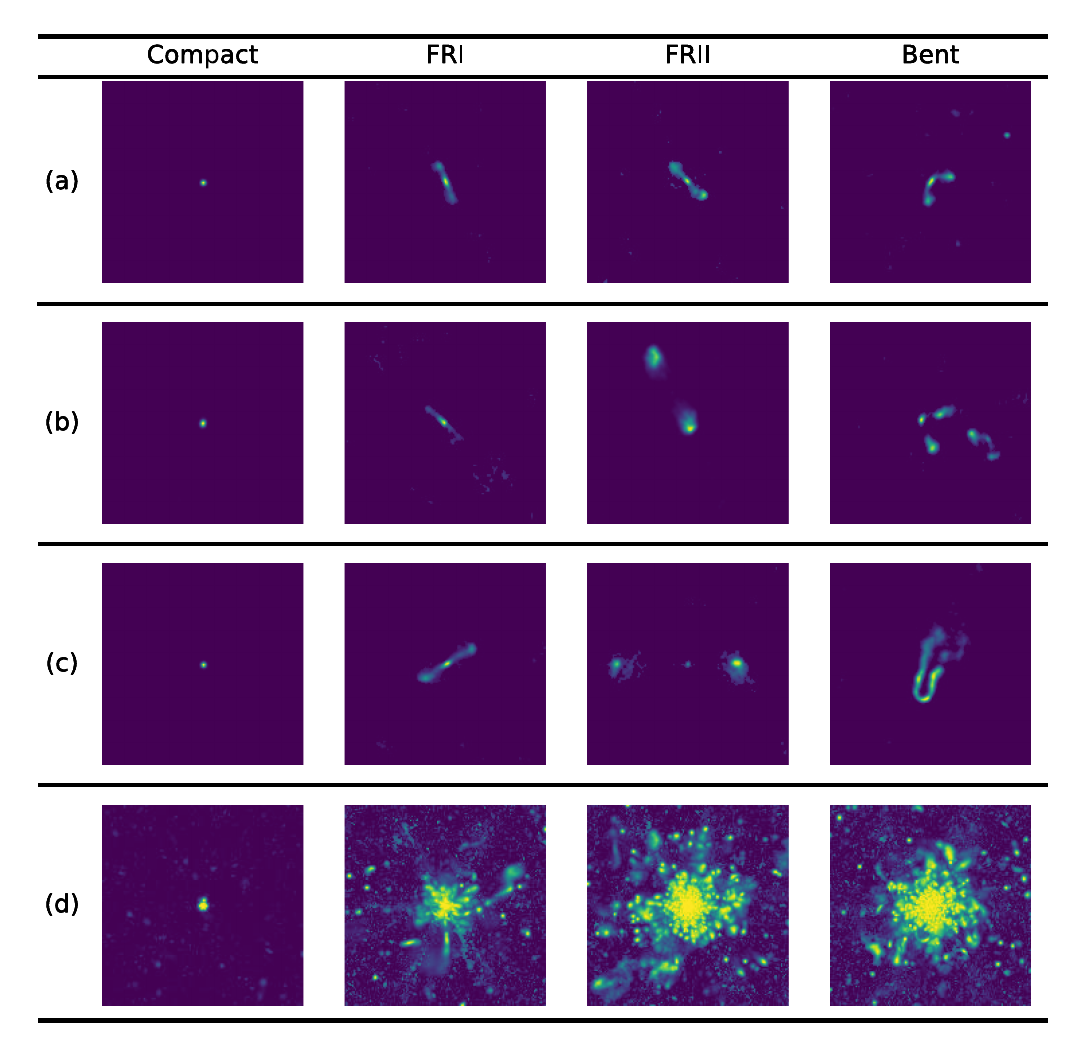}
\caption{(a-c) Sample illustration of the Compact, FRI, FRII and Bent radio sources morphological appearance. (d) Maximum image superposition of all Compact, FRI, FRII and Bent radio galaxies from the data set used in this paper.}
\label{fig:max_images_per_class}
\end{figure}

Radio galaxies can be divided into four primary categories based on their intrinsic morphological differences. The most important two are Fanaroff \& Riley I (FRI) and Fanaroff \& Riley II (FRII) \citep{fanaroff1974morphology}. FRI galaxies have bright cores relative to their lobes (have decreasing luminosity from the core). FRII galaxies, on the other hand, have edge-brightened lobes with a core at the centre. The third and most common category of radio galaxies is the Compact class, which refers to point-like radio sources \citep{Baldi2015APS,baldi2018fr0cat}. The fourth category is Bent, which is composed of radio sources with jets that are bent at an angle, either in a narrow-angled tail (NAT) or a wide-angled tail (WAT) configuration \citep{rudnick1976head}. While the Compact and FRI classes are quite distinctive, the FRII and Bent classes are much more challenging. In Fig.~\ref{fig:max_images_per_class} we illustrate a few examples of each of the four classes as well as superimposed variants created by utilizing all images from the data set described in Table~\ref{tab:original_dataset_distrb}. In addition to the difficulties related to their shape characteristics, the classification of FRI, FRII, and Bent galaxies is further exacerbated by the fact that the galaxies in the data set have differing orientations. Upon visual inspection, it becomes apparent that each category exhibits a discernibly distinct brightness distribution. Nonetheless, it is worth noting that the FRII and Bent classes share substantial similarities, making it difficult to discern between the two classes hence a challenging task to differentiate them.

In this paper, we propose a lightweight paradigm that involves trainable COSFIRE (Combination of Shifted Filter Responses) filters \citep{azzopardi2012trainable,azzopardi2016gender}, which have been found effective in various computer vision applications. This approach is efficient, learning-free, rotation-tolerant, explainable, and does not require a huge training set. 

The implementation of a COSFIRE-based classification pipeline is relatively easy and straightforward from a conceptual standpoint as described in Section \ref{sec: methods}. 
It involves the configuration of COSFIRE filters \citep{azzopardi2012trainable} whose selectivity of each filter is automatically determined from the shape properties of a single training example. The objective is to set up multiple filters, whose combined responses generate a feature signature for the type of galaxy present in an image. This approach is analogous to how visual cells in the mammalian brain encode visual information, a concept known as population coding \citep{pasupathy1999responses,pasupathy2002population}. COSFIRE filters have been applied in various computer vision tasks: gender recognition \citep{azzopardi2016gender}, contour detection \citep{azzopardi2012corf} and handwritten digit classification \citep{azzopardi2013shape}. In this work, COSFIRE filters are configured to extract the hyperlocal geometric arrangements that uniquely describe the patterns of radio sources (in terms of blobs) in a given image. 

The rest of the paper is structured as follows. Section \ref{sec: relatedworks} presents the current state-of-the-art approaches. Section \ref{sec: data} describes the data set used in this study. Section \ref{sec: methods} describes the proposed COSFIRE-based paradigm. Section \ref{sec: evaluation} presents the evaluation criteria used to assess the performance of our approach. Section \ref{sec: experiments_and_results} describes the experiments and the results obtained. Section \ref{sec: discussion} provides a discussion of the results obtained in relation to the relevant work. Finally, we draw conclusions in Section \ref{sec: conclusion}.

\section{Related Works}
\label{sec: relatedworks}
Different automated computer intelligence paradigms have been exploited to automate the rather daunting manual classification approaches that radio astronomers use to identify different types of radio sources \citep{ndungu2023advances}. Convolutional neural networks (CNNs) have dominated the field in recent years. In particular, \cite{aniyan2017classifying}, used the AlexNet architecture \citep{krizhevsky2017imagenet}, calling the trained model Toothless\footnote{https://github.com/ratt-ru/toothless}, to achieve accuracies of 91\%, 75\% and 95\% for the FRI,  FRII, and Bent-tailed morphologies, respectively. Subsequently, notable incremental breakthroughs have been made in the applications of deep learning to the field of radio astronomy, ranging from shallow CNN architectures \citep{Lukic_2019} to deep and complex architectures such as DenseNet \citep{huang2017densely,samudre2022data}. Furthermore, other noteworthy advancements within the field are: model-centric strategies such as group-equivariant CNNs (G-CNNs) \citep{scaife2021fanaroff} to support equivariance translations on various isometries of radio galaxies and multidomain multibranch CNNs \citep{tang2022radio} to allow models to learn jointly from various survey inputs; data-centric approaches such as data augmentation \citep{maslej2021morphological,kummer2022radio,ma2019machine,slijepcevic2022radio}; transfer learning \citep{2019MNRAS.488.3358T,Lukic_2019} and $N$-shot learning\footnote{The algorithms for $N$-shot learning have been developed to optimally utilize the limited amount of supervised information that is accessible, i.e., labelled data sets, to create precise predictions while circumventing the obstacles like overfitting.} \citep{samudre2022data} to overcome the limited availability of annotated data sets in radio astronomy. These models have been shown to perform competitively, providing promising alternatives to prior models such as the one by \citet{aniyan2017classifying}, which showed signs of overfitting.

Conventional machine learning has also been explored for the morphological classification of Fanaroff-Riley (FR) radio galaxies \citep{2021AJ....161...94S,becker2021cnn,10.1093/mnras/stab271,darya2023morphological}. This paradigm involves extracting handcrafted features such as intensity, shape, lobe size, number of lobes, and texture from input radio image data, which are then utilized in machine learning algorithms \citep{brand2023feature}. Previous studies have leveraged feature descriptors to capture the texture of radio images, such as Haralick features, which consist of a group of 13 non-parametric measures derived from the Grey Level Co-occurrence Matrix \citep{10.1093/mnras/stab271}. Gradient boosting methods \citep{FRIEDMAN2002367} such as XGBoost \citep{chen2016xgboost}, LightGBM \citep{ke2017lightgbm} and CatBoost \citep{dorogush2018catboost} have been explored \citep{darya2023morphological}. Research has shown that feature engineering can provide machine learning algorithms with highly significant features, leading to promising results \citep{2021AJ....161...94S}. Moreover, \citet{darya2023morphological} demonstrated that conventional machine learning (LightGBM,  XGBoost, and CatBoost) perform competitively to CNN-based models with deep learning on relatively small data sets (about 10,000 images and below). 

CNNs are regarded as the state-of-the-art in various image classification applications \citep{aniyan2017classifying, Lukic_2019, scaife2021fanaroff, tang2022radio}. However, they require large amounts of training data and are susceptible to overfitting when trained with small data sets, which is the case in radio astronomy. Moreover, the high computational demands of deep architectures for training and applying CNNs often require GPUs, which can be costly and limit their applicability in resource-limited settings. Moreover, CNN-based models lack insufficient intrinsic robustness to rotations. To address rotational variations in radio sources, multiple approaches have been taken. One approach is to utilize group-equivariant CNNs, where the network is designed to capture the diverse orientation information of a given input galaxy in encoded form \citep{scaife2021fanaroff}. Another method involves augmenting the training data by applying rotations to the training samples, enabling the CNNs to learn different orientations of the classes. Furthermore, a pre-processing step can be employed to standardize the rotation of all radio sources. This may be achieved by using principal component analysis (PCA) to align the galaxies' principal components with the coordinate system's axes, effectively normalizing their orientations \citep{brand2023feature}. 

As evident from this literature review, numerous challenges remain, including the need for efficient (computationally inexpensive) and rotationally invariant methods. In this work we address these limitations with the proposed COSFIRE filter approach.

\section{Data}

\label{sec: data}
The data set of radio galaxies used in this paper was compiled and processed by \citet{samudre2022data}. It was constructed by selecting well-resolved radio galaxies from multiple catalogs: Proctor catalog \citep{proctor2011morphological} (derived from the FIRST\footnote{Faint Images of the Radio Sky at Twenty Centimeters} survey, 2003) for the Bent radio galaxies; FR0CAT \citep{baldi2018fr0cat} and CoNFIG\footnote{Combined NVSS–FIRST galaxies} catalogs \citep{gendre2008combined,10.1111/j.1365-2966.2010.16413.x} for Compact radio galaxies; FRICAT \citep{capetti2017fricat} and CoNFIG catalogs for FRI radio galaxies and finally FRIICAT catalog  \citep{capetti2017friicat} and CoNFIG catalogs for FRII radio galaxies. This data set is used in this paper with the objective of conducting comparative analyses of similar work done by \citet{samudre2022data}.

The initial data set is composed of the following classes: Compact (406 samples), Bent (508 samples), FRI (389 samples), and FRII (679 samples). These samples are further divided into training, validation, and testing data sets as shown in Table~\ref{tab:original_dataset_distrb}. According to \citet{samudre2022data}, the original data set's underrepresented classes were balanced by adding randomly duplicated samples to the training and validation data sets. Table~\ref{tab:balanced_dataset_distrb} depicts the distribution of the balanced data set.

The images were pre-processed by utilizing sigma-clipping with a threshold of 3$\sigma$ \citep{Aniyan_2017}. This technique involves eliminating or discarding pixels that have background noise levels above or below 3 standard deviations from the mean \citep{Aniyan_2017}. 

Although there exist more recent catalogs as described by \citet{ndungu2023advances}, the latest catalogs, such as LoTSS (DR1 \& DR2)\footnote{The LOFAR Two-metre Sky Survey (Data Release I \& II)} \citep{shimwell2019lofar,shimwell2022lofar}, are not annotated into Compact, FRI, FRII and Bent classes. Instead, they are labelled by automated tools such as the Python Blob Detector and Source-Finder (PyBDSF) \citep{mohan2015pybdsf} that categorize astronomical sources into three types: `S' for single isolated sources modeled with one Gaussian, `C' for sources that are within a group but can be individually modeled with a single Gaussian, and `M' for extended sources that need multiple Gaussians for accurate modeling. This system aids in the efficient identification and analysis of space emissions.

\begin{table}
\footnotesize
\centering
\caption{\textbf{The size of the \textit{original} data set distributed across the training, validation, and test categories.}}
\label{tab:original_dataset_distrb}
\begin{tabular}{@{}p{2.4cm}p{0.7cm}p{0.7cm}p{0.9cm}p{1cm}p{0.7cm}@{}}
\toprule
\textbf{Source catalog} & \textbf{Type} & \textbf{Total} & \textbf{Training} & \textbf{Validation} & \textbf{Test} \\
\midrule 
Proctor & Bent & 508 & 305 & 100 & 103 \\
FR0CAT \& CoNFIG & Compact & 406 & 226 & 80 & 100 \\
FRICAT \& CoNFIG & FRI & 389 & 215 & 74 & 100 \\
FRIICAT \& CoNFIG& FRII & 679 & 434 & 144 & 101 \\
\midrule 
Total & & 1982 & 1180 & 398 & 404 \\
\bottomrule
\end{tabular}
\end{table}

\begin{table}
\centering
\footnotesize
\caption{\textbf{The size of the \textit{balanced} data set distributed across the training, validation, and test categories.}}
\label{tab:balanced_dataset_distrb}
\begin{tabular}{@{}p{2.4cm}p{0.7cm}p{0.7cm}p{0.9cm}p{1cm}p{0.7cm}@{}}
\toprule
\textbf{Source catalog} & \textbf{Type} & \textbf{Total} & \textbf{Training} & \textbf{Validation} & \textbf{Test} \\
\midrule 
Proctor & Bent & 680 & 433 & 144 & 103 \\
FR0CAT \& CoNFIG & Compact & 675 & 431 & 144 & 100 \\
FRICAT \& CoNFIG & FRI & 674 & 430 & 144 & 100 \\
FRIICAT \& CoNFIG& FRII & 679 & 434 & 144 & 101 \\
\midrule 
Total & & 2708 & 1728 & 576 & 404 \\
\bottomrule
\end{tabular}
\end{table}

\section{Methods}
\label{sec: methods}

\begin{figure*}
 \centering
 \footnotesize
\includegraphics{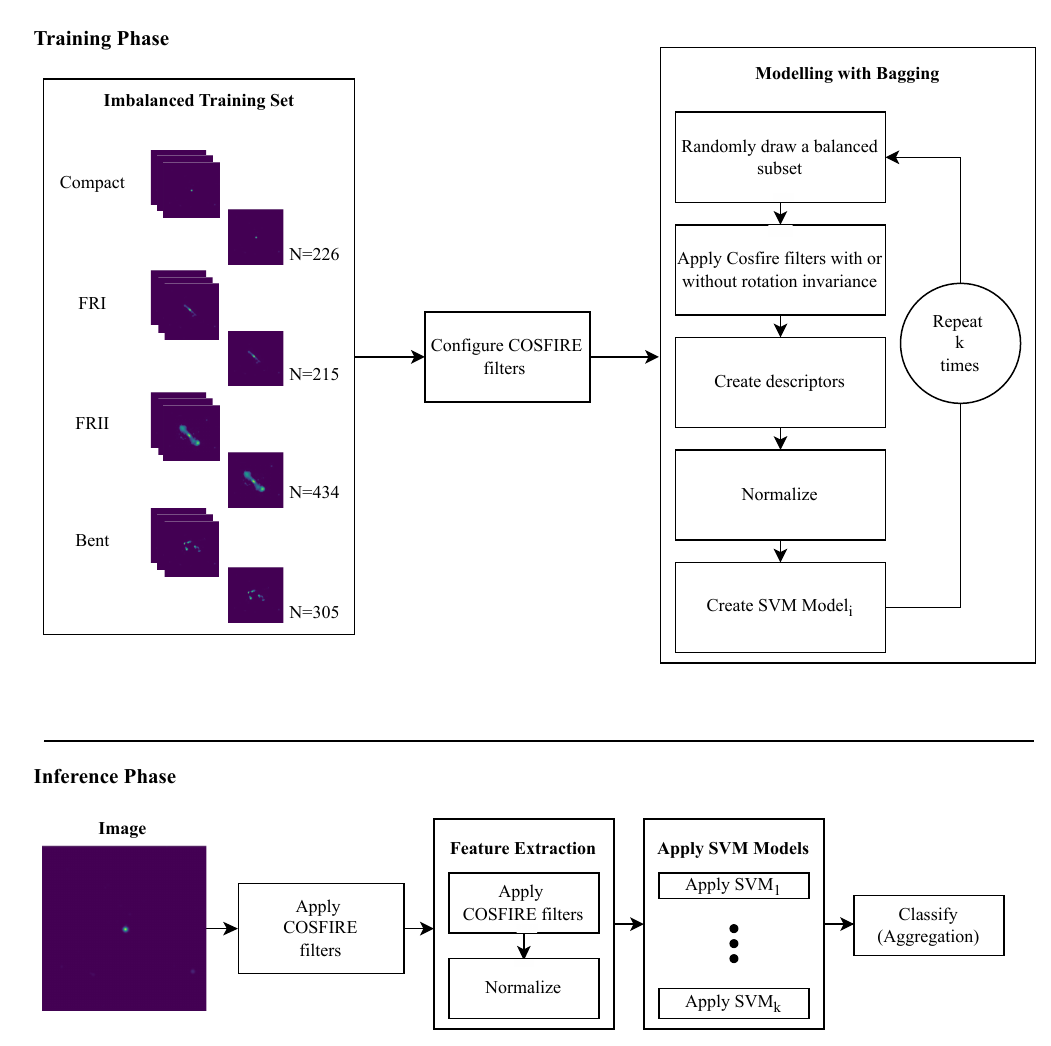}
\caption{A graphic representation of the proposed COSFIRE filter pipeline for radio galaxy classification for both training and inference phases. The training phase involves configuring COSFIRE filters, extracting descriptors and training an ensemble of classifiers using a bagging approach. These models are applied in the inference stage and the label of the given image is determined by decision fusion.}
\label{fig:methodology_figure}
\end{figure*}

 This section gives a comprehensive overview of the methodology that we propose for radio galaxy image classification with the COSFIRE filter approach. We explain the process of radio source blob detection, the configuration of COSFIRE filters with rotation invariance, the formation of feature descriptors, and the utilization of these descriptors for classification. An overview of this pipeline is illustrated in Fig. \ref{fig:methodology_figure}.
 
\subsection{Blob detection}
\label{sec:blob_detector}
Blob detection is a technique used for identifying points or regions in an image that exhibit a sudden change in intensity (areas that are either brighter or darker than the surrounding areas), known as a "blob". This approach enables the identification of regions that may correspond to objects or structures of interest within the image in our case radio source(s). In computer vision, one of the commonly used blob detectors is based on the Laplacian of Gaussian (LoG) \citep{wang2017blob,wang2020automated}. The LoG is the second-order derivative of the Gaussian function, denoted by $G(x,y)$:

\begin{equation}
G(x, y ; \sigma)=\frac{1}{2 \pi \sigma^{2}} \exp \left(-\frac{x^{2}+y^{2}}{2 \sigma^{2}}\right)
\end{equation}
\noindent where $\sigma$ is the standard derivation. 
The Laplacian of Gaussian is typically estimated by a Difference-of-Gaussian (DoG) function, which is separable, and thus convolving it with a 2D image is much more computationally efficient. Given two Gaussian functions, $G_1$ and $G_2$ with their respective standard deviations $\sigma_1$ and $\sigma_2$, the Gaussians are defined as:

\begin{equation}
G_{1}(x,y;\sigma_1) = \frac{1}{2\pi \sigma_1^2} \exp\bigg(-\frac{x^2 + y^2}{2\sigma_1^2}\bigg)
\end{equation}

\begin{equation}
G_{2}(x,y;\sigma_2) = \frac{1}{2\pi \sigma_2^2} \exp\bigg(-\frac{x^2 + y^2}{2\sigma_2^2}\bigg)
\end{equation}

\noindent The DoG function is derived by subtracting two Gaussian functions with different standard deviations:

\begin{equation}
DoG(x,y;\sigma_1,\sigma_2) = G_{1}(x,y;\sigma_1) - G_{2}(x,y;\sigma_2)
\label{eq:DOG1}
\end{equation}

For a point $(x, y)$ and an image $I$ with intensity distribution $I\left(x^{\prime}, y^{\prime}\right)$, we calculate the response $c_{\sigma}(x, y)$ of a DoG filter with a kernel function $\operatorname{DoG}_{\sigma}\left(x-x^{\prime}, y-y^{\prime}\right)$ by convolution:

\begin{equation}
c_{\sigma_1,\sigma_2}(x, y) = ReLU(I \star Do G_{\sigma_1,\sigma_2})_{t_1}
\label{eq:convolution}
\end{equation}

\noindent where the rectification linear unit (ReLU) is an activation function that sets to zero all values below $t_1$ and $\star$ represents convolution.

When $\sigma_1<\sigma_2$ we refer to the resulting $DoG$ function as a center-on $DoG$ or $DoG^+$ for brevity, with the central region exhibiting a positive response and the surround exhibiting a negative response. Conversely, when $\sigma_2 < \sigma_1$ the configuration results in a center-off $DoG$ function, which we denote by $DoG^-$. In this work, we adopt the approach used by \citet{azzopardi2015trainable} and always set the smaller standard deviation to be half of the larger standard deviation.

\begin{figure}
 \centering
 \footnotesize
\includegraphics[width=\columnwidth]{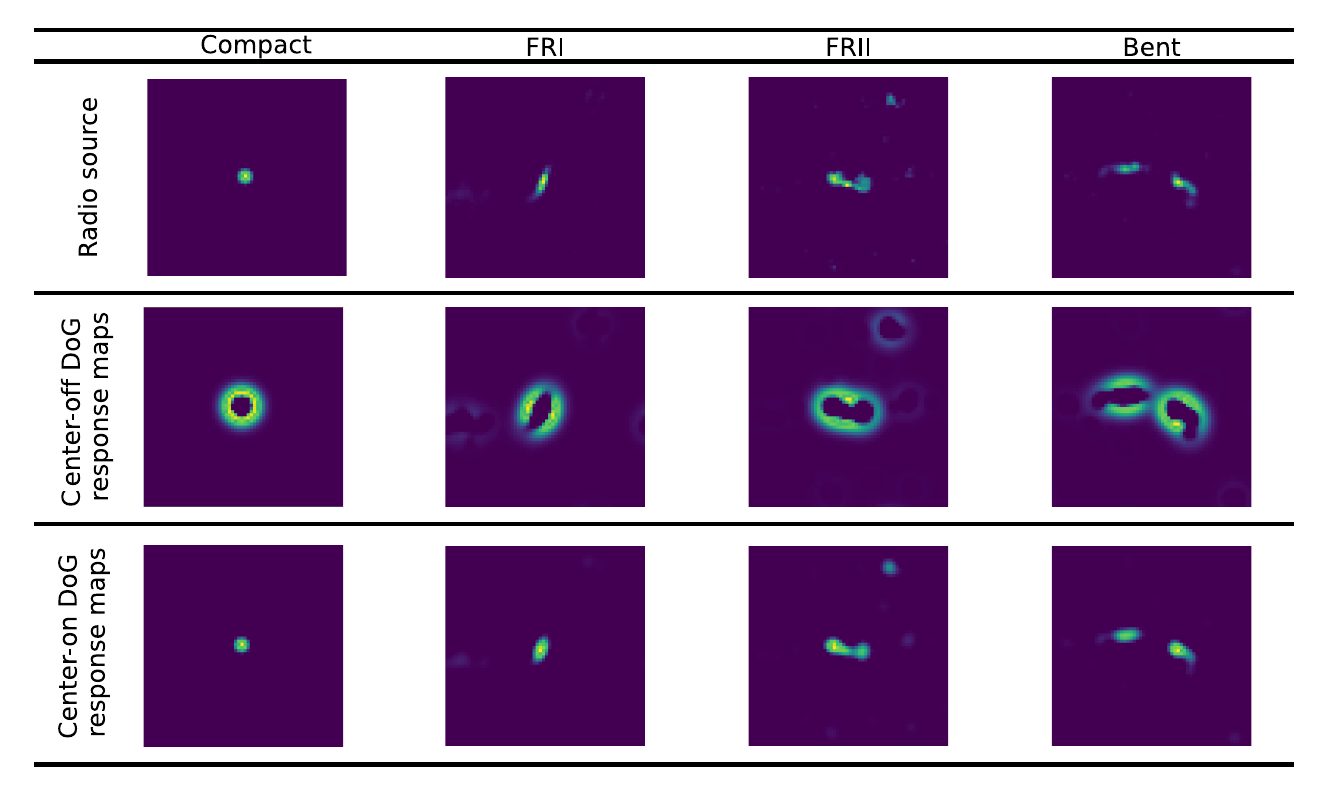}

\caption{Centre-on and centre-off DoG filter response maps obtained with convolution on Compact, FRI, FRII and Bent radio sources utilizing a standard deviation $\sigma$ of 3 for the outer Gaussian function. The images are of size 150 $\times$ 150 pixels.}
\label{fig:DoG_Responses_res.svg}
\end{figure}

The centre-on DoG filter highlights bright blobs on a dark background and is more sensitive to intensity increases at the centre of the blob. This filter responds well to objects and is less sensitive to edges or sharp changes in intensity. On the other hand, the centre-off DoG filter highlights dark blobs on a bright background and is more sensitive to intensity decreases at the centre of the blob. This filter responds well to corners and edges and is less sensitive to objects or regions of uniform intensity. Therefore,  both centre-on and centre-off DoG filters are used to detect blobs or regions of interest in an image, see Fig. \ref{fig:DoG_Responses_res.svg} for an illustration on Compact, FRI, FRII and Bent radio sources. The DoG filter is a type of band-pass filter that can eliminate high-frequency components that represent noise as well as some low-frequency components that represent homogenous areas.

\subsection{COSFIRE filter configuration}
\label{sub_sect:COSFIREfilterconfiguration}

A COSFIRE filter is automatically configured by examining the shape properties of a given prototype pattern of interest in an image, which ultimately determines its selectivity. 
The process of configuration can be summarized in three main steps: \textit{convolve-ReLU-keypoint detection}. The first two steps involve the convolution of center-on and center-off DoG filters followed by ReLU as described above. Finally, keypoint detection requires the determination of local maximum thresholded DoG responses along a set of concentric circles around a point of interest. For this application, we use the center of the image as the point of interest, but in principle, any location can be used for this purpose. The point of interest is the location that best characterizes a pattern of interest. The number and radii of the concentric circles along with the threshold $t_1$ used by the ReLU function are hyperparameters of the COSFIRE approach. 

A keypoint denoted by $k$, which is identified as a local maximum of a DoG filter along a concentric circle, is characterized by a four-element tuple: ($\sigma_{k}$, $\delta_{k}$, $\rho_{k}$, $\phi_{k}$). Here, $\sigma_k$ and $\delta_k$ indicate the standard deviation of the outer Gaussian function of the DoG function and its polarity that achieved the highest response in the polar coordinates with radius $\rho_k$ at an angle of $\phi_k$ radians with respect to the given point of interest. We denote by $C_f$ a COSFIRE filter, which is represented as a list of such tuples:

\begin{equation}
 C_{f} = \{(\sigma_{k},\delta_k,\rho_{k},\phi_{k}) ~|~ k=1,\dots, n \}
	\label{eq:tuples}
\end{equation}

\noindent where $n$ refers to the number of DoG responses considered in $C_f$, which plays a crucial role in the selectivity and generalization of the COSFIRE filter. Typically, selectivity increases and generalization decreases with increasing value of $n$ (i.e. number of keypoints).

\begin{figure}
    \centering
    \footnotesize
    \begin{tabular}{cc}
        \includegraphics[width=3.8cm, height=3.8cm]{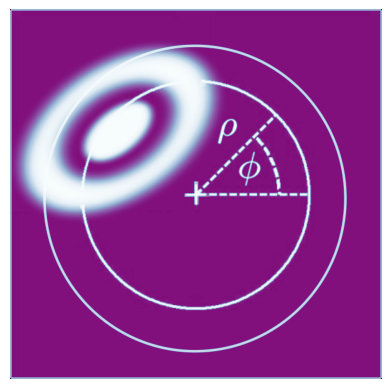} &
        \includegraphics[width=3.8cm, height=3.8cm]{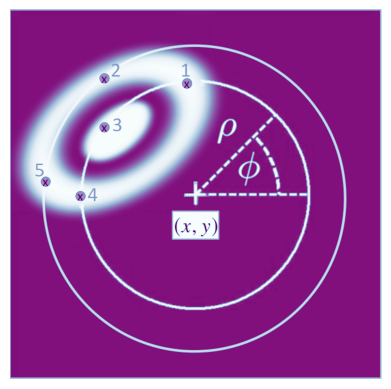} \\
        (a) & (b) \\
    \end{tabular}   
    \caption{Example of keypoint detection for the configuration of a COSFIRE filter. (a) Synthetic image simulating the superposition of a center-on and a center-off DoG response maps. (b) The five black dots indicate the five keypoints that are identified in locations that exhibit local maximum DoG values along the given two radii $\rho_{k_1}$ and  $\rho_{k_2}$ at an angle $\phi_{k_i}$ with respect to the center $(x,y)$.}
    \label{fig:cosfire_config}
\end{figure}

Fig.~\ref{fig:cosfire_config} illustrates an example of the automatic configuration of a COSFIRE filter from the superposition of synthetic center-on and center-off DoG maps. The example uses two concentric circles and the resulting COSFIRE filter is a set of five tuples describing the five keypoints indicated in Fig.~\ref{fig:cosfire_config}b. The keypoints are identified at the positions along the circles where the DoG responses reach local maxima. 

\subsection{COSFIRE filter response}
\label{sub_sect:COSFIRE_filter_response}
The response of a COSFIRE filter $C_f$ for a given $(x,y)$ location is computed by combining the responses of the DoG functions whose scale $\sigma_k$, polarity $\delta_k$ and position ($\rho_k,\phi_k)$ with respect to $(x,y)$ are indicated in the set $C_f$. The process of computing the response can be summarized in five main steps: \textit{convolve-ReLU-blur-shift-combine}. The \textit{convolve} step refers to the convolution of center-on and center-off DoG filters followed by the \textit{ReLU} operation that sets to zero all values below the given threshold $t_1$. These are the same two steps that were required for the configuration described above. Then, in order to allow for some tolerance in the preferred positions of the involved DoG responses, we \textit{blur} each DoG response with a Gaussian function $G_{\sigma^{\prime}}(x^{\prime}, y^{\prime})$, whose standard deviation $\sigma^{\prime}$ is a linear function of the distance $\rho_k$: $\sigma^{\prime} = \lfloor\sigma_0^{\prime} + \alpha\rho_k\rfloor$, with $\sigma_0^\prime$ and $\alpha$ being hyperparameters. Moreover, we \textit{shift} the blurred responses in the direction opposite of the polar coordinates such that all DoG responses of interest meet at the same $(x,y)$ location. Finally, the \textit{combine} step is implemented as suggested by \cite{azzopardi2012trainable}, where a COSFIRE filter response, which we denote by $r_{C_f}$, is obtained by the geometric mean function that {combines} all blurred and shifted thresholded DoG responses:

\begin{equation}
r_{C_{f}} (x,y)=\left(\prod_{k=1}^{n} z_{\sigma_{k}, \delta_k, \rho_{k}, \phi_{k}}(x, y)\right)^{\frac{1}{n}}
	\label{eq:cosfire_response}
\end{equation}

\noindent where 
\begin{equation}
z_{\sigma_{k}, \delta_k, \rho_{k}, \phi_{k}}(x, y)=\max _{x^{\prime}, y^{\prime}}\{c_{\sigma_{k}}(x-\Delta x_{k}-x^{\prime}, y-\Delta y_{k}-y^{\prime})G_{\sigma^{\prime}}(x^{\prime}, y^{\prime})\}
\end{equation}

\noindent is the combined blurred and shifted DoG filter response map of tuple $k$. The shifting operation displaces each blurred DoG response of interest to the support centre of the COSFIRE filter. The shift vector is denoted by $\left(\Delta x_{k}, \Delta y_{k}\right)$, where $\Delta x_{k}=-\rho_{k} \cos \phi_{k}$ and $\Delta y_{k}=-\rho_{k} \sin \phi_{k}$. Additionally, $-3 \sigma^{\prime} \leqslant x^{\prime}, y^{\prime} \leqslant 3 \sigma^{\prime}$.

\subsection{Tolerance to rotations}
Tolerance to rotations can be attained by configuring multiple COSFIRE filters using rotated versions of a single prototype pattern. An effective approach to achieve this involves defining new filters by modifying the parameters of an existing COSFIRE filter. For instance, consider a COSFIRE filter denoted as $C_{f_\psi}$, designed to be selective for the same underlying pattern that was employed to configure the original COSFIRE filter $C_f$, but rotated by $\psi$ radians. This new filter is defined as follows:

\begin{equation}
C_{f_\psi} = \{(\sigma_{k}, \delta_k, \rho_{k}, \phi_{k}+ \psi)~|~\forall~ (\sigma_{k}, \delta_k, \rho_{k}, \phi_{k}) \in C_f \}
	\label{eq:rotation_tuples}
\end{equation}

The rotation-tolerant response of a COSFIRE filter $\hat{r}_{C_f}$ is then achieved by taking the maximum response across all COSFIRE filters selective for the same pattern at 12 equally-spaced orientations:

\begin{equation}
    \hat{r}_{C_f}(x,y) = \max \bigg \{r_{C_{f_\psi}(x,y)} ~\forall~\psi \in \{0, \pi/6, \dots, 11\pi/6\} \bigg\}
\end{equation}

\subsection{COSFIRE descriptor}
\label{subsec: descriptors}
For a given image, a COSFIRE descriptor, which we denote by $D$ and define below, is generated by applying all COSFIRE filters in rotation-tolerant mode and extracting the maximum value from each filter, regardless of its location. Consequently, for a set of $\lambda$ filters, the resulting description of a given image is represented by a $\lambda$-dimensional vector.

\begin{equation}
    D = \bigg[\max_{x,y}\bigg\{\hat{r}_{C_{f_1}}(x,y)\bigg\}, \dots,\max_{x,y}\bigg\{\hat{r}_{C_{f_\lambda}}(x,y)\bigg\}\bigg]
\end{equation}

This concept is inspired by neurophysiology research, which suggests that the shape representation of a stimulus is based on the combined activity of a group of shape-selective neurons in area V4\footnote{V4 cells are neurons in the visual cortex that are involved in form perception (recognizing objects and their features such as shape).} \citep{Wielaard01howsimple, azzopardi2012corf,weiner2015population}.

\subsection{COSFIRE descriptor pre-processing}

The only pre-processing step done to the COSFIRE descriptors is L2 normalization\footnote{The L2 norm is determined by taking the square root of the sum of the squares of all the values in the vector. Also known as Euclidean norm.}. Data normalization is a common preprocessing step in tabular machine learning tasks that transforms the input data into a \emph{more uniform} scale across the features. In particular, the Support Vector Machines (SVM) objective function includes the distance between data points, making it sensitive to the scale of the input features. Therefore, non-normalized features with larger scales can dominate distance calculations and may hence affect the performance \citep{hsu2003practical}.

\subsection{Classification model}

The SVM model, introduced by \citet{cortes1995support}, has been selected for the classification of COSFIRE-based image descriptors due to its capacity to handle high-dimensional data, manage outliers, and achieve robust generalization. The training of the SVM model is conducted using the COSFIRE descriptors extracted from a training set that encompasses four distinct classes of radio galaxies. Given that the training set is characterized by an imbalance in class distribution (as indicated in Table~\ref{tab:original_dataset_distrb}), a bagging (bootstrap aggregating) approach is employed to train an ensemble of classification models. This technique improves model generalization and suppresses any biases towards the majority classes \citep{breiman1996bagging}. Figure~\ref{fig:methodology_figure} illustrates the entire process of training and inference. In the training phase, we employ a resampling technique with replacement to randomly select a balanced subset of 290 images from each class. The rationale behind this choice is as follows: we aim to create a balanced training data set for learning an SVM model. To achieve this balance, we calculate two-thirds of the size of the majority class (in this case, 434), which results in 290 images. By selecting 290 images from each class, we also ensure that the resulting balanced subset is roughly the same size as the original imbalanced training set. With the subsets, we then train a set of ten different SVM models. In the inference phase, we use the ten SVM models to perform classification by aggregating their predictions on the test data set. In practice, the subset size and the number of classifiers are two hyperparameters, which can be fine-tuned by cross validation. However, we choose to work with the mentioned default values as the fine-tuning of such parameters is beyond the scope of the proposed COSFIRE paradigm.

\section{Performance metrics for evaluation}
\label{sec: evaluation}
\subsection{Accuracy}
As in prior research on morphological classification  \citep{Lukic_2019, samudre2022data}, we evaluate the performance of the proposed COSFIRE approach following the widely adopted accuracy metric. To assess the performance of our model in the experiments, we leverage the validation set for determining the hyperparameter values.

\subsection{Floating Point Operations (FLOPs)} 

FLOPs is a measure of computer performance that indicates the number of floating-point operations that a processor executes to complete a task. This measurement is commonly used in scientific programs/applications that heavily depend on floating-point (FP) calculations such as CNNs. We will evaluate the computational complexity of our approach by measuring the number of floating point operations (FLOPs) required, and then compare it to the computational complexity of DenseNet161 \citep{samudre2022data}. 

FLOPs estimation for both algorithms is performed during the inference phase rather than the training phase of the classification process. This phase entails only the forward pass of input data through the model to produce predictions, focusing on the computational cost of executing the already trained model.


\section{Experiments and results}
\label{sec: experiments_and_results}
In this section, we present and analyze the experimental results obtained in our research on radio galaxy classification. We also compare the performance and computational complexities of our method with other existing approaches in radio galaxy classification.

\subsection{Performance}
We conducted a series of experiments to evaluate the performance of the proposed trainable COSFIRE filter approach. As detailed in Section \ref{sec: methods}, our pipeline involves the automatic configuration of COSFIRE filters and their application to training, validation and test data sets for each distinct class (Bent, Compact, FRI, and FRII). Our primary objective was to attain the highest classification accuracy for distinguishing among the four radio source classes. Through this rigorous evaluation process, we conducted an in-depth exploration of the trainable COSFIRE filters' performance and their potential in the realm of radio source classification.

We used the validation data set to determine the hyperparameters of the COSFIRE filter configuration ($\sigma$, $P$\footnote{These are the radii used to configure COSFIRE filters.}, $t_1$\footnote{This is the threshold used by the ReLU function.}), and application  ($\sigma_{0}^\prime$, and $\alpha$). In our experimentation, we explored three values per hyperparameter: $\sigma \in \{5,6,7\}$, 
$P \in \{\{0,5,10,15,20\}, \{0,5,10,15,20,25\}, \{0,5,10,15,20,25,30\}\}$,
$t_1 \in \{0.05,0.1,0.15\}$, $\sigma_0^\prime \in \{0.5,0.75,1\}$, and $\alpha \in \{0.1,0.15,0.2\}$. This resulted in a total of 243 unique parameter sets. For every unique set, we configured up to 100 COSFIRE filters for each class, leading to a total of 400 COSFIRE filters. The filters were configured by selecting random images from the training set. To compensate for the randomness of the image selection, we executed three experiments with the same set of hyperparameters and then took the average results across the three experiments. These experiments resulted in a $243\times400$ matrix of accuracy rates. We then identified the maximum accuracy rate for each row (i.e. for each set of hyperparameters). Importantly, multiple sets of hyperparameters yielded very close accuracies. Therefore, to account for all the hyperparameters generating similar results, we performed a right-tailed student $t$-test statistic between the row with the global maximum accuracy rate and all the other rows. It turned out that 26 sets of hyperparameters yielded results that are not statistically different from the global maximum. The average maximum accuracy of the experiments with these 26 sets of hyperparameters was achieved with 93 COSFIRE filters per class on the validation set.

\begin{figure}
 \centering
 \footnotesize

\includegraphics[width=8.5cm]{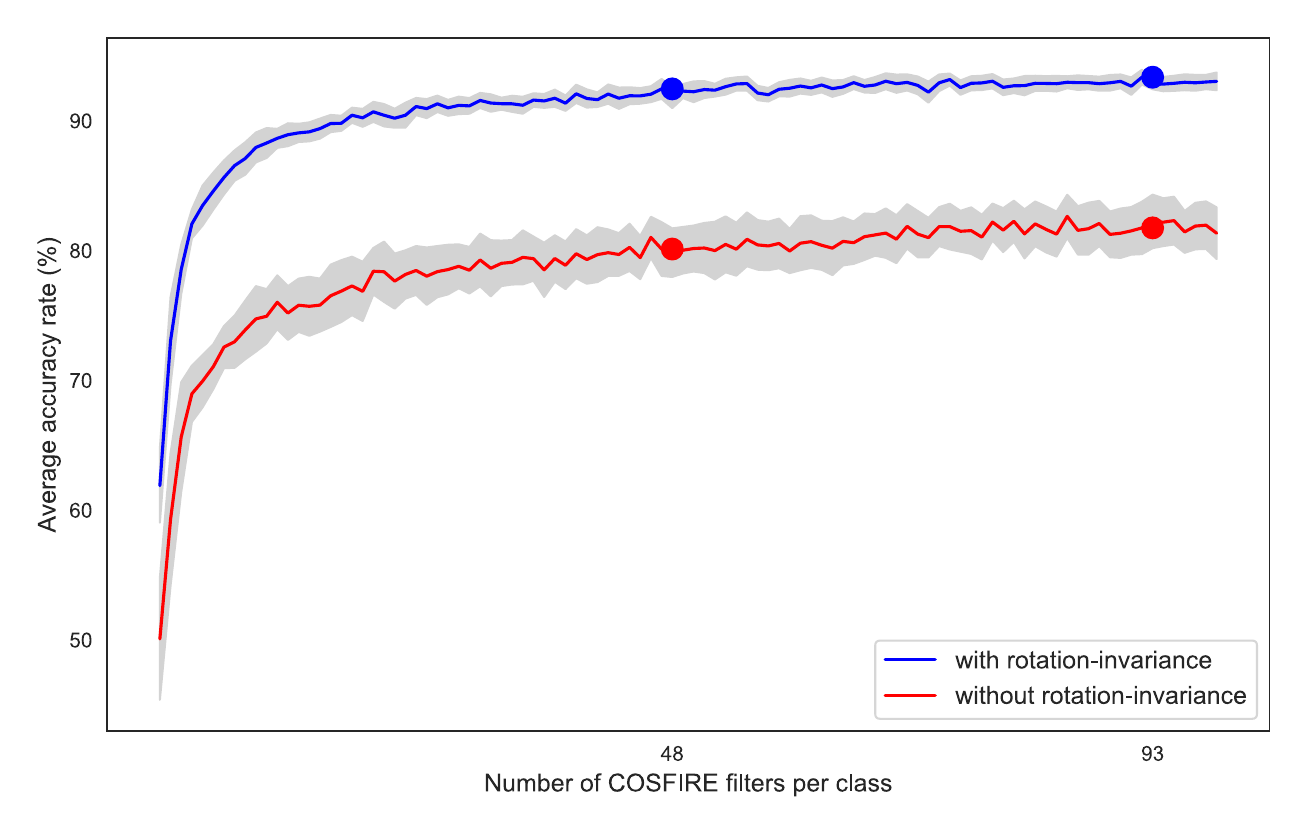}

\caption{Accuracy rate achieved on the original test set (see Table~\ref{tab:original_dataset_distrb}) as a function of the number of COSFIRE filters used. The plots demonstrate how the average accuracy rate over the experiments with the determined 26 sets of hyperparameters varies based on the number of COSFIRE filters used, considering both cases of rotation invariance and without rotation invariance. Specifically, employing 48 filters for each class approximates the 92\% accuracy level of DenseNet161, as reported by \citet{samudre2022data}. Using 93 filters per class, which led to the maximum average accuracy on the validation set, we achieve an accuracy of 93.36\% on the test set.}
\label{fig:accuracy_vs_cosfire_filter}
\end{figure}

Subsequently, we employ the COSFIRE filters in conjunction with the corresponding classifiers, which are configured using 26 distinct sets of hyperparameters determined previously on the validation set, to the test data. The outcomes of this process are graphically represented in Fig.~\ref{fig:accuracy_vs_cosfire_filter}. This figure includes two principal plots illustrating the variation in accuracy rates as a function of the number of COSFIRE filters used. One plot delineates the performance when the filters operate in a rotation-invariant manner, while the other depicts the scenario without rotation invariance. Each plot is the average across the results obtained by the experiments using the 26 sets of hyperparameters, with the grey shading indicating the standard deviations. With 93 filters per class (as determined from the validation data) and rotation invariance, the average accuracy on the test set is 93.36\% ± 0.57 (with the minimum being 92.24\% and the maximum being 94.31\%), whereas, without rotation invariance, the same filters yielded an average accuracy of 82.25\% ± 2.06. This performance trend is visually depicted in Fig.~\ref{fig:accuracy_vs_cosfire_filter}, highlighting that as the number of filters increases, the disparity in performance between including and excluding rotation becomes more pronounced. Additionally, the figure shows that the model performance is more stable with rotation invariance than without it. The latter has high variability even with more COSFIRE filters.

\subsection{Comparison with previous works}

The most direct comparison to our study is the research that used DenseNet-161 \citep{huang2017densely} and Siamese networks \citep{koch2015siamese} methodologies conducted by \citet{samudre2022data} on the same data. In their work, they conducted their experiments on two versions of the data: the original and the balanced data set as presented in Table~\ref{tab:original_dataset_distrb} and Table~\ref{tab:balanced_dataset_distrb}, respectively. According to their reported results, Siamese networks and DenseNet-161 achieved the highest accuracies of 71.1\% ± 0.40  and 91.2\% ± 0.60, respectively, on the original data set. On the other hand, Siamese networks achieved 73.9\% ± 0.50  and DenseNet-161 achieved 92.1\% ± 0.40 accuracy on the balanced data set. Notably, our COSFIRE approach achieves an accuracy of 93.36\% ± 0.57 without data augmentation or any further image pre-processing. This is more than 16\% reduction in the classification error rate compared to the DenseNet-161 model \citep{samudre2022data}. The significant reduction in the error rate is mainly attributed to the rotation invariance properties of the COSFIRE approach. We did not perform runs on the balanced data set (Table~\ref{tab:balanced_dataset_distrb}) since this would not generate new insights, as the data set was augmented through image duplication of the minority classes.

\subsection{Computational complexity}
\subsubsection{Convolutional Neural Networks}
Calculating the FLOPs in a  convolutional neural network involves determining the number of arithmetic operations performed during the forward pass. This includes the multiplications and additions between the input data elements and the trainable convolution kernels. The FLOPs computation takes into account factors such as the input size, kernel size, number of filters, and the presence of padding, dilation, and stride. A CNN convolutional operation with padding set to 0, dilation set to 1, and stride set to 1, for simplicity \citep{freire2022computational}, can be summarised as follows:  

\begin{equation}
y_{x, y}^{f} = \phi\left(\sum_{i=1}^{n_{H}}\sum_{j=1}^{n_{W}} \sum_{k=1}^{n_{C}}K_{i,j,k}^{f} \cdot I_{x+i-1, y+j-1, k}   + b^{f}\right)
\label{eq:cnn_rm}
\end{equation}

\noindent In equation \ref{eq:cnn_rm}, the output of a specific convolutional layer is denoted by $y_{x, y}^{f}$, which represents a feature map. Here, the superscript $f$ indicates the filter index, while the subscripts $x$, $y$, and $k$ represent the row, column, and channel indices of the output feature map, respectively. The trainable convolution kernel is denoted by $K_{i, j, k}^{f}$, and it performs element-wise multiplications with the corresponding input elements of image $I$ of dimensions $n_H\times n_W\times n_C$\footnote{To meet the input requirements of a DenseNet model, the original images are resized from $150\times150$ to $224\times224\times3$ pixels.}. The bias term $b^{f}$ is added to the sum, and the resulting value undergoes the activation function $\phi(\cdot)$ to generate the output feature map element $y_{x, y}^{f}$. This equation embodies the convolution operation in a CNN, where each output feature map element is computed by convolving the corresponding region of the input feature maps with trainable filters and introducing non-linearity through the activation function.

The output size of the CNN at each layer is important when calculating the FLOPs of the CNN architecture used. It is given by,

\begin{equation}
\text {OutputSize} =\left[\frac{n_{h}+2p - k}{\text {s}}+1\right],
\label{eq:output_size}
\end{equation}

\noindent where $n_h$ is the image input size, $p$ is padding, $k$ is the kernel size and $s$ is the stride. The number of FLOPs can then be estimated as:

\begin{equation}
\text{FLOPs}_{\text{CNN}}=n_{f}n_{i} n_{k}  \cdot \text{OutputSize}
\label{eq:flops}
\end{equation}

\noindent where $n_{f}$ is the number of filters. In each sliding window, there are $n_{i}$ $n_{k}$ multiplications and the sliding window process needs to be repeated as many times as the \text { OutputSize }. This entire procedure is then repeated for all $n_{f}$ filters. The estimated number of FLOPs of the DenseNet-161 model is 15.6 GFLOPs based on the workflow\footnote{Roughly GMACs = 0.5 * GFLOPs} \footnote{https://github.com/sovrasov/flops-counter.pytorch} of \citep{ptflops}. Importantly, the 15.6 GFLOPs calculations include the operations involving the dense and activation layers in the DenseNet architecture.

\subsubsection{COSFIRE filter approach}
The FLOPs associated with the single convolution required by the COSFIRE filter approach are computationally inexpensive due to the exploitation of the separability property of Gaussian filters. For the separability of the Gaussian filter, equation \ref{eqn: gaus_sep}, the 2D Gaussian can be written as the multiplication of two functions: one that depends on $x$ and the other that depends on $y$. In this case, the two functions are the same and they are both 1D Gaussian\footnote{This reduces the number of operations required to apply the filter to an image. For example, applying a $5\times5$ filter to an image requires 25 multiplications per pixel, but applying two 1D filters of length 5 requires only 10 multiplications per pixel.}. Because the DoG function combines two Gaussian functions linearly, the separability property is also preserved in the DoG function.

\begin{equation}
\begin{aligned} G(x, y ; \sigma) & =\frac{1}{{2 \pi \sigma^{2}}} \exp \left(-\frac{x^{2}+y^{2}}{2 \sigma^{2}}\right) \\ & =\frac{1}{\sqrt{2 \pi} \sigma}  \exp \left(-\frac{x^{2}}{2 \sigma^{2}}\right) \times \frac{1}{\sqrt{2 \pi} \sigma} \exp \left(-\frac{y^{2}}{2 \sigma^{2}}\right)
\end{aligned}
\label{eqn: gaus_sep}
\end{equation}

\begin{figure}
 \centering
 \footnotesize

\includegraphics[width=8.5cm]{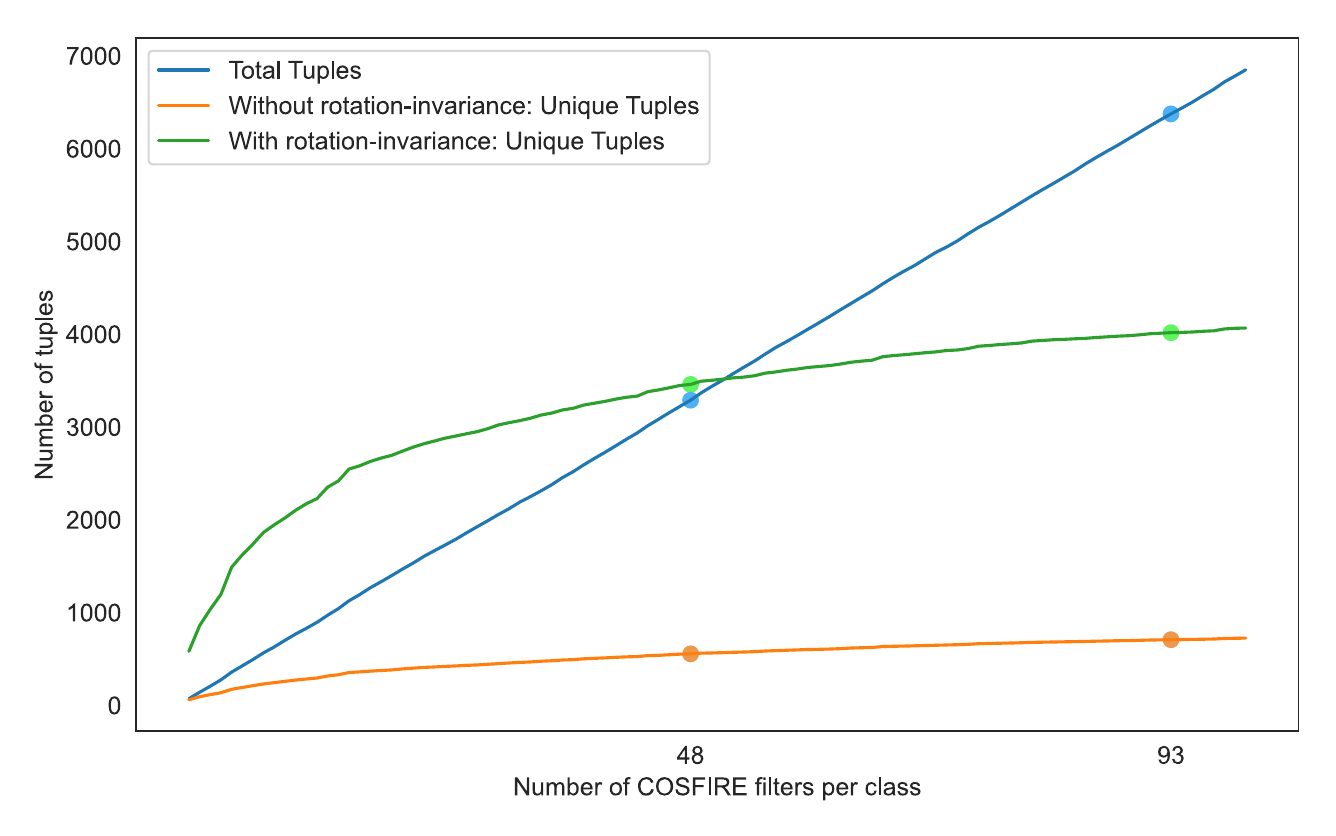}

\caption{Number of tuples as a function of the number of COSFIRE filters used per class. The total number of tuples increases linearly with the increase in the number of COSFIRE filters per class, but the number of unique tuples increases sublinearly, since most tuples are shared among all COSFIRE filters and stored in the memory, hence requiring less computations. The total number of tuples required by 48 filters is very close to the number required by 93 filters.}
\label{fig:cosfire_operators_plot}
\end{figure}

Table \ref{tab:flop_estimation} provides a breakdown of the FLOPs that are required by the COSFIRE methodology at the inference stage. In that Table, we use the following set of hyperparameters as an example: $\sigma=5, P \in \{0,5,10,15,20\}$, $t_1=0.1$, $\sigma_0=0.5$, and $\alpha=0.1$. The FLOPs computations are mainly on the processes of convolve-ReLU-blur-shift-combine. For a given image of $150\times150$ pixels and 400 COSFIRE filters, the total estimated number of FLOPs is $\sim$1.5 GFLOPs. Also, in order to optimize the computational efficiency in a system using 4800 (i.e. 100 filters for each of the four classes applied in 12 orientations) COSFIRE filters that collectively involve 64,284 tuples, we employ the following strategy:

\begin{itemize}
\item \textbf{Elimination of Redundancy}: Recognizing that numerous tuples are repeated across these filters, we first isolate every distinct tuple from the entire set. In fact, the number of tuples increases sublinearly with the increase in the number of COSFIRE filters used, Fig.~\ref{fig:cosfire_operators_plot}. 
\item \textbf{Computation and Storage}: We calculate the response map for each unique tuple only once. This response map is essentially a modified version of the center-on or center-off DoG response map, subjected to blurring and shifting. We then store each computed response map in a hash table for quick retrieval, avoiding redundant computations.
\item \textbf{Configuration Sharing}: Upon further analysis, we find that the configurations of many COSFIRE filters share identical pairs of tuples.
\item \textbf{Pairwise Response Maps}: We compile a list of these common tuple pairs and pre-calculate the combined response map for each pair by performing a multiplication of the individual response maps.
\item \textbf{HashTable for Pairs}: The resulting combined response maps of tuple pairs are also stored in a hash table. This ensures that the response for any pair that is used by more than one filter can be quickly fetched without recalculating it.
\end{itemize}
By implementing this approach, we make the process more efficient, as we avoid unnecessary recalculations for both individual tuples and their pairs, which significantly reduce the computational load and speeds up the overall filtering process. Using this efficient technique, the total number of FLOPs required is reduced to $\sim$1.1 GFLOPs; i.e. 26\% reduction in the FLOPs (comparison of FLOPs basic and FLOPs column values in Table \ref{tab:flop_estimation}). 

\begin{figure}
 \centering
 \footnotesize
\includegraphics[scale=0.38]{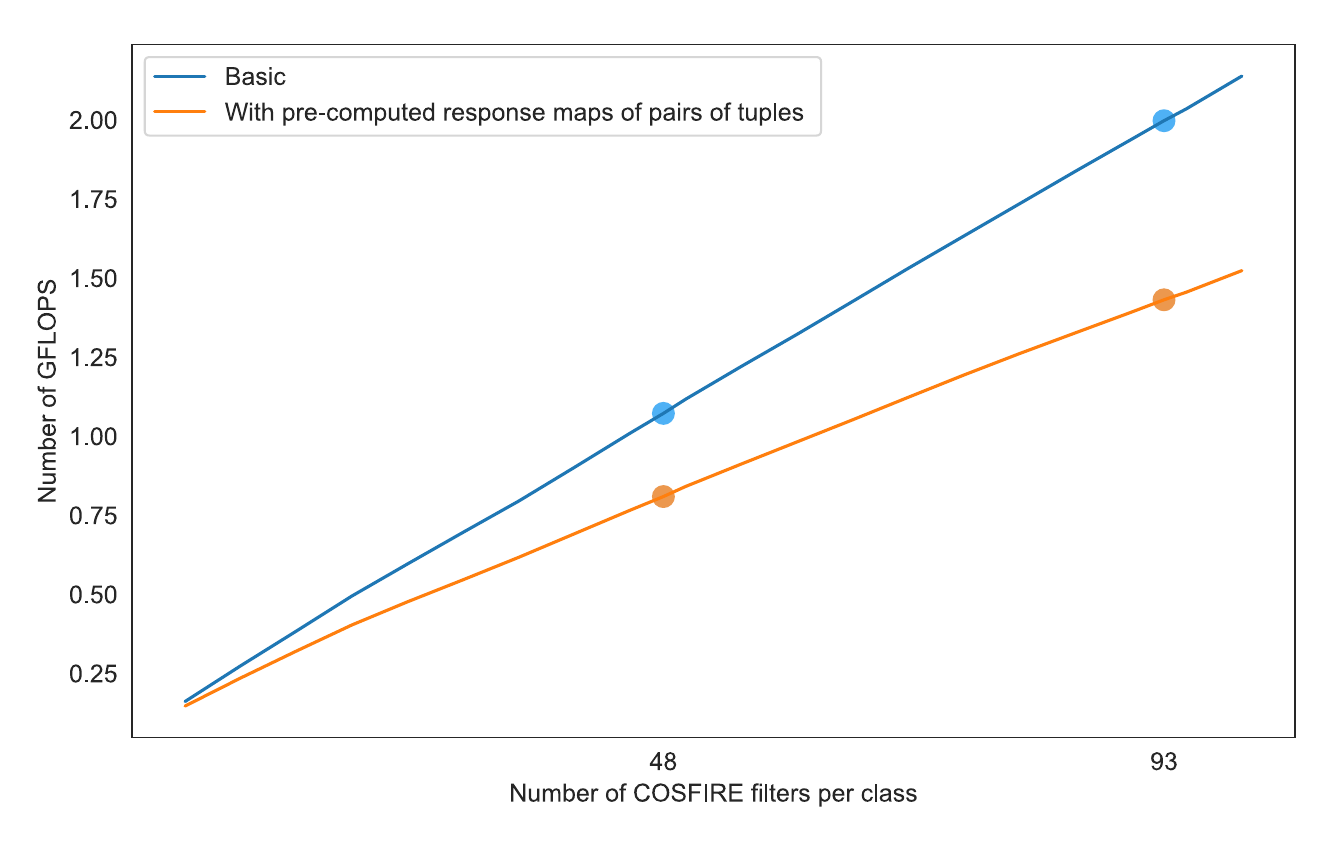}
\caption{The number of FLOPs with and without the pre-computation of response maps of pairs of tuples against the number of COSFIRE filters per class. }
\label{fig:cosfire_flop_plot}
\end{figure}

Similarly, conducting three trials with 26 sets of optimal hyperparameters from the validation data set, we noted a significant decrease in FLOPs, as depicted in Figure~\ref{fig:cosfire_flop_plot}, corresponding to an increasing number of COSFIRE filters. In fact, with reference to Fig.~\ref{fig:cosfire_flop_plot} and Fig.~\ref{fig:accuracy_vs_cosfire_filter} we demonstrate that we can achieve very high performance with fewer filters using the COSFIRE approach. Using just 48 filters for each class yields an accuracy rate of 92.46\% ± 0.76, with a computational cost of 0.8 GFLOPs—approximately 20 times less than the DenseNet161 model that achieves a similar result.

\renewcommand{\arraystretch}{1.5}
\begin{table*}
\centering
\footnotesize
\caption{\textbf{COSFIRE filter FLOPs estimation at inference stage. The FLOPs calculation in the columns \emph{FLOPs basic} and \emph{FLOPs} (computation with pre-computed response maps of pairs of tuples) are based on the number of tuples obtained when using a single image for inference. As an example here is a breakdown of FLOPs at each stage for the specific set of hyperparameters: $\sigma=5, P \in \{0,5,10,15,20\}$, $t_1=0.1$, $\sigma_0^\prime=0.5$, and $\alpha=0.1$. In the \emph{FLOPs} column, the symbol -- represents the equivalent value from the \emph{FLOPs basic} column.}}
\label{tab:flop_estimation}
\begin{tabular}{@{}p{8.9cm}p{4.1cm}p{1.7cm}c@{}}
\toprule
\textbf{Step} & \textbf{Formula} & \textbf{FLOPs basic} & \textbf{FLOPs}  \\
\midrule 
\textbf{Center-On DoG sigma:} Convolve the $m \times n (150\times 150)$ image with a DoG kernel whose size is $s=6\sigma+1$ (here $\sigma=5$). The expression $a$ combines the product and summation operations, and the result is multiplied by 2 to account for the vertical and horizontal separable filters of the DoG functions. &  $a = 2(m n(2s-1))$ & 2,745,000  & -- \\
\textbf{Center-off DoG (same sigma):} Center-off is simply the negative of the center-on response map. & $b= mn$ & 22,500 & --  \\
\textbf{ReLU of center-on DoG map:} Set all values below $t_1$ to zero across the center-on responses using ReLU activation function. & $c= mn$ & 22,500 &  -- \\
\textbf{ReLU of center-off DoG map:} Same as above but for the center-off response map. & $d= mn$ & 22,500 &  --  \\
\multirow[t]{5}{8.9cm}{\justifying \textbf{Blurring:} Similar to $a$, convolve the $m \times n~(150\times 150)$ image with a DoG kernel of size: ${\sigma^\prime}_{k} = \lfloor\sigma_0 + \alpha\rho_k\rfloor$. The expression $a$ is multiplied by two to account for center-on and center-off DoG response maps. The extent of blurring depends on the  radii: 0, 5, 10, 15, and 20. The kernel size of the blurring function for the $k$-th radius is calculated as $s_k = 6\hat{\sigma}_{k}+1$: $s_0=6(1)+1=7,~s_5=7,~s_{10}=13,~s_{15}=13,~s_{20}=19$.}& 
\begin{tabular}{@{}p{1cm}@{}l@{}p{2.75cm}@{}}$\hat{e}_{(\rho_k=0)}$ & ~=~ & $2(2(m n(2(7)-1)))$\\ \end{tabular} & 1,170,000 & --\\ [-6pt]
& \begin{tabular}{@{}p{1cm}@{}l@{}p{2.75cm}@{}}$\hat{e}_{(\rho_k=5)}$ & ~=~ & $2(2(m n(2(7)-1)))$\\ \end{tabular} & 1,170,000 & --\\ [-6pt]
& \begin{tabular}{@{}p{1cm}@{}l@{}p{2.75cm}@{}}$\hat{e}_{(\rho_k=10)}$ & ~=~ & $2(2(m n(2(13)-1)))$ \end{tabular}& 2,250,000 & --\\ [-6pt]
& \begin{tabular}{@{}p{1cm}@{}l@{}p{2.75cm}@{}}$\hat{e}_{(\rho_k=15)}$ & ~=~ & $2(2(m n(2(13)-1)))$ \end{tabular}& 2,250,000 & --\\ [-6pt]
& \begin{tabular}{@{}p{1cm}@{}l@{}p{2.75cm}@{}}$\hat{e}_{(\rho_k=20)}$ & ~=~ & $2(2(m n(2(19)-1)))$ \end{tabular}& 3,330,000 & --\\ [-3pt]
& $e=\sum_{i \in \{0,5,10,15,20\}} \hat{e}_{(\rho_k=i)}$ \\
\textbf{Shifting:} Shift the two DoG response maps as many times as the number of all unique tuples $t_p$ across all COSFIRE filters. In our experiments, the $t_p$ was 3187  (with rotation invariance). & $f = mnt_p$ & 71,707,500 &  -- \\
\textbf{Multiplication:}  Compute the responses of all COSFIRE filters by multiplying the corresponding shifted response maps. The number of rotations is denoted by $n_r=12$. The variable $n_d=4800$ represents the number of COSFIRE filters, which here is set to 400 (100 per class) and multiplied by the total number of rotations ($n_r=12$). $T_p$ represents the total number of tuples across all the COSFIRE filters ($T_p= 64,284$). \newline The set of total tuples configured considering rotation invariance have repetitions that do not need to be re-computed. In this step, we pair the tuples that appear more than once among the 64,284 tuples. Then, we only compute and store in the memory a single pair of the shared tuples (duplicates) to save on computations. Therefore, the FLOPS are calculated from the pre-computed response maps of pairs of tuples and those of unique tuples.  &  $g = mnT_p - mnn_d$ &  1,338,390,000
  & 937,305,300 \\
\textbf{Hashkey:} Before applying the multiplication operation, the shifted response map must first be retrieved from a hashtable residing in memory. The keys of this hashtable are determined by multiplying the first four prime numbers raised to given $\sigma$, $\delta$, $\rho$ and $\phi$: $h_k=2^\sigma3^\delta5^\rho7^\phi$. The retrieval of each tuple response map from memory is therefore 7 FLOPS.   & $h=T_ph_kn_r$. & 449,988
 &  -- \\
\textbf{Taking the Root for Geometric Mean:} Geometric mean calculation operations for each response map of all $n_d$ COSFIRE filters.  & $i= mnn_dn_r$ & 108,000,000  &  -- \\
\textbf{Descriptor formation:}  Determination of a $n_{dp}=400$-element descriptor by taking the maximum value of each of the $n_{dp}$ response maps. & $j = mnn_{dp}$ & 9,000,000  &  --  \\
\textbf{Decision fusion of 10 SVM classifiers:} SVM calculations depend on the number of decision function hyperplanes ($n_{hp}$) given the number of classes under classification. This experiment uses the one-vs-one (`ovo') decision function approach, which means that $n_{hp} = z(z-1)/2$. In this case, $z=4$ and hence $n_{hp}=6$. Each hyperplane separates one class from the rest of the classes. The number of FLOPs is therefore based on the dot products needed for one SVM between the given feature vector of size $n_{dp}$ and the $n_{hp}$ hyperplanes. Since the dot product involves multiplications and additions then one SVM takes $2n_{dp}n_{hp}$ FLOPs. Having 10 SVM, this is repeated 10 times and finally, we choose the maximum which takes another 9 FLOPs. &  $k= 10(2n_{dp}n_{hp}$) + 9 & 48,009 &  -- \\
\midrule 
\textbf{Total FLOPs} & $  a+b+c+d+e+f+g+h+i+j+k$ & 1,540,825,497 & 1,139,740,797 \\ 
\bottomrule
\end{tabular}
\end{table*}

\section{Discussion}
\label{sec: discussion}

The proposed trainable COSFIRE filter approach, which is designed to capture radio galaxies at different orientations, achieved better results than the CNN-based approach for radio galaxy classification on the same data set. This suggests that the trainable COSFIRE filter approach can capture more relevant salient features of the radio galaxy images than the CNN-based approach. Rotation invariance plays a vital role in robust galaxy classification\footnote{The shape of radio galaxies that we observe depends on how their jets are oriented in the plane of the sky.}, facilitating accurate and consistent outcomes. Also, the limited availability of annotated training data is attributed to the high cost of the labelling and the limited number of experienced astronomers dedicated to this task (considering the size of the astronomical data generated by modern telescopes). Therefore, for effective model generalization, it is essential that a classification model remains invariant to rotations, ensuring consistent predictions irrespective of the input's orientation.

Our results are consistent with previous studies exploring different approaches to incorporating rotation invariants of radio galaxies in the training set as a means of improving classification model accuracy and generalization. In other similar studies, in radio astronomy, \citet{brand2023feature} showed that classification performance is improved when the orientations of training galaxies are normalized as opposed to when no such attempt was made to address rotational variations. \citet{scaife2021fanaroff} used group-equivariant convolutional neural networks that encode multiple orientation information of a given input galaxy. Moreover, the widely adopted data augmentation pre-processing step in machine model development is another approach that also addresses this challenge of radio galaxy equivariance \citep{becker2021cnn}. While data augmentation is a useful tool, especially in small data sets, it comes with its own problems, namely the potential introduction of unrealistic or irrelevant variations into the data, which can lead to model overfitting and reduced interpretability. In addition, it requires domain knowledge to develop a good augmentation strategy, which considers all possible equivariant transformations given a galaxy image. The rotation invariance in the COSFIRE approach is intrinsic to the method and does not require data augmentation. This makes the approach more robust, versatile, and completely data-driven, with no domain expertise required, and hence highly adaptable to various computer vision applications.

CNN-based networks are computationally expensive. For instance, the forward pass of DenseNet161 consumes $\sim$15.6 GFLOPs. On the other hand, the COSFIRE filter approach demonstrates efficiency by utilizing a substantially lower number of FLOPs (Fig.~\ref{fig:cosfire_flop_plot}), resulting in significantly lower demand for computational resources. The workflow of the COSFIRE approach is also more flexible than conventional CNN architectures. The number of filters used is a hyperparameter of this paradigm, which is in contrast to the fixed number of filters used in its counterpart.  Our approach, based on the COSFIRE filter, is a novel algorithm for the classification of radio galaxies and comes with a low computational cost. 

\section{Conclusion}
\label{sec: conclusion}
In this work, we introduce a novel descriptor based on trainable COSFIRE filters for radio galaxy classification. We combine this descriptor with an SVM classifier using a linear kernel and achieve superior results compared to a DenseNet161 transfer learning-based pre-trained network and a few-shot learning-based Siamese network on the same training data set. Our technique is computationally inexpensive ($\sim$20$\times$ lower cost with the same accuracy as the DenseNet161 model), rotation invariant, free from data augmentation and does not rely on domain expert knowledge. These features make our technique not only effective (in terms of accuracy) but also efficient (in terms of FLOPs) for radio galaxy image classification tasks. The accuracy rate of 93.36\% that we achieved is the best result ever reported for the benchmark data set prepared by \citep{samudre2022data}\footnote{\url{https://github.com/kiryteo/RG\_Classification\_code}}.

This work contributes to the field of radio astronomy by providing an alternative technique for identifying and analyzing radio sources. As the next-generation telescopes (such as LOFAR, MeerKAT, and SKA) produce high-resolution images of the sky, our future work will explore the effectiveness of our methodology on these new data and assess the possibility of cross-survey predictions.

\section*{Acknowledgements}

\noindent Part of this work is supported by the Foundation for Dutch Scientific Research Institutes. This work is based on the research supported in part by the National Research Foundation of South Africa (grant numbers 119488 and CSRP2204224243).

\noindent We thank the Center for Information Technology of the University of Groningen for their support and for providing access to the Hábrók high performance computing cluster.

\section*{Data Availability}

The data set is available at: \url{https://github.com/kiryteo/RG\_Classification\_code}.



\bibliographystyle{mnras}
\bibliography{bibliography}







\bsp	
\label{lastpage}
\end{document}